\documentclass[12pt]{article}

\newread\testifexists
\def\GetIfExists #1 {\immediate\openin\testifexists=#1
    \ifeof\testifexists\immediate\closein\testifexists\else
    \immediate\closein\testifexists\input #1\fi}

\usepackage{gthstyle}
\immediate\openin\testifexists=e
    \ifeof\testifexists\immediate\closein\testifexists\else
    \immediate\closein\testifexists\input e\fipsf
\def\Bbb#1{\setbox0=\hbox{$\tt #1$}  \copy0\kern-\wd0\kern .1em\copy0}
\def\bbf#1{\setbox0=\hbox{$#1$} \kern-.025em\copy0\kern-\wd0
        \kern.05em\copy0\kern-\wd0 \kern-.025em\raise.0433em\box0}
\immediate\openin\testifexists=a
    \ifeof\testifexists\immediate\closein\testifexists\else
    \immediate\closein\testifexists\input a\fimssym.def          %% for \Bbb A - Z %%

\def\a{\alpha}      \def\b{\beta}         
\def\d{\delta}      \def\D{\Delta}  \def\e{\varepsilon}
\def\et{\eta}               
                     \def\vv{\varphi}
\def\n{\nu}         \def\j{\psi}    
\def\r{\varrho}     \def\s{\sigma}  
\def\t{\tau}        \def\th{\theta}  
              
\def\w{\omega}        

  \def\OO{{\cal O}}
\def\pa{\partial} \def\ra{\rightarrow}

\def\dd{{\rm d}}     \def\ket{\rangle}

\def\deff{\ {\buildrel{\rm def}\over{=}}\ }

\def\fract#1#2{{\textstyle{#1\over#2}}}
\def\ffract#1#2{\raise .3 em\hbox{$\scriptstyle#1$}\kern-.25em/
                \kern-.2em\lower .2 em \hbox{$\scriptstyle#2$}}

\def\half{\fract12}  

\def\part#1#2{{\partial#1\over\partial#2}}

\newcommand{\be}{\begin{eqnarray}}
\renewcommand{\le}[1]{\label{#1}\end{eqnarray}}
\newcommand{\ee}{\end{eqnarray}}
\newcommand{\eqn}[1]{(\ref{#1})}

\newcommand{\nm}{\nonumber}

\newcommand{\fn}{\footnote}
\newcommand{\newsec}[1]{\section{#1}\setcounter{equation}{0}}
\begin{document}

\begin{titlepage}
\begin{center}
\hfill ITP-UU-02/69  \\ \hfill SPIN-2002/45  \\ \hfill {\tt
quant-ph/0212095}\\ \vskip 20mm

{\large \bf DETERMINISM BENEATH QUANTUM MECHANICS\fn{Presented at
``Quo vadis Quantum Mechanics?", Temple University, Philadelphia,
September 25, 2002.}.}

\vskip 10mm

{\large\bf Gerard 't~Hooft}

\vskip 4mm Institute for Theoretical Physics \\  Utrecht
University, Leuvenlaan 4\\ 3584 CC Utrecht, the Netherlands
\medskip \\ and
\medskip \\ Spinoza Institute \\ Postbox 80.195 \\ 3508 TD
Utrecht, the Netherlands
\smallskip \\ e-mail: \tt g.thooft@phys.uu.nl \\ internet:
\tt http://www.phys.uu.nl/\~{}thooft/

\vskip 6mm

\end{center}

\vskip .2in
        %\begin{center} {\bf Astract } \end{center}
\begin{quotation} \noindent {\large\bf Abstract } \medskip \\
Contrary to common belief, it is not difficult to construct
deterministic models where stochastic behavior is correctly
described by quantum mechanical amplitudes, in precise accordance
with the Copenhagen-Bohr-Bohm doctrine. What is difficult however
is to obtain a Hamiltonian that is bounded from below, and whose
ground state is a vacuum that exhibits complicated vacuum
fluctuations, as in the real world.

Beneath Quantum Mechanics, there may be a deterministic theory
with (local) information loss. This may lead to a sufficiently
complex vacuum state, and to an apparent non-locality in the
relation between the deterministic (``ontological") states and
the quantum states, of the kind needed to explain away the Bell
inequalities.

Theories of this kind would not only be appealing from a
philosophical point of view, but may also be essential for
understanding causality at Planckian distance scales.
\end{quotation}

\vfill \flushleft{\today}

%%%%%%%%%%%%%%%%%%

\end{titlepage}

\eject

%%%%%%%%%%%%%%%%%%%%%%%%%%%%%%%%%%%%%%%%%%%%%%%%%%%%
\newsec{Motivation}
The need for an improved understanding of what Quantum Mechanics
really is, needs hardly be explained in this meeting. My primary
concern is that Quantum Mechanics, in its present state, appears
to be mysterious. It should always be the scientists' aim to take
away the mystery of things. It is my suspicion that there should
exist a quite logical explanation for the fact that we need to
describe probabilities in this world quantum mechanically. This
explanation presumably can be found in the fabric of the Laws of
Physics at the Planck scale. % Sherlock Holmes

However, if our only problem with Quantum Mechanics were our
desire to demystify it, then one could bring forward that, as it
stands, Quantum Mechanics works impeccably. It predicts the
outcome of any conceivable experiment, apart from some random
ingredient. This randomness is perfect. There never has been any
indication that there would be any way to predict where in its
quantum probability curve an event will actually be detected. Why
not be at peace with this situation?

One answer to this is Quantum Gravity. Attempts to reconcile
General Relativity with Quantum Mechanics lead to a jungle of
complexity that is difficult or impossible to interpret
physically. In a combined theory, we no longer see ``states" that
evolve with ``time", we do not know how to identify the vacuum
state, and so on. What we need instead is a unique theory that
not only accounts for Quantum Mechanics together with General
Relativity, but also explains for us how matter behaves. We
should find indications pointing towards the correct unifying
theory underlying the Standard Model, towards explanations of the
presumed occurrence of supersymmetry, as well as the mechanism(s)
that break it. We suspect that deeper insights in what and why
Quantum Mechanics is, should help us further to understand these
issues.

Related to the question of quantizing gravity is the problem of
quantizing cosmology. Astrophysicists tell us that the Universe
started with a ``big bang", but, at least at first sight, such a
statement appears to be at odds with the notions of quantum
mechanical uncertainty. In principle, we could know the state the
Universe is in presently, and then one could solve the
Schr\"odinger equation backwards in time, but this should yield a
quantum superposition of many configurations, not just a Big
Bang. Questions of this sort may seem of purely academic nature,
but they become very concrete as soon as one attempts to
construct some reasonable model for a ``Quantum Universe". The
notion of a quantum state of the Universe appears to defy logic.

Attempts nevertheless to reconcile Quantum Mechanics with
Cosmology were made. Whether the proposed schemes may be viewed as
a satisfactory picture of our world, is difficult to discuss. To
convince someone that they are flawed may be as difficult as
changing someone's religious beliefs. Therefore, I shall refrain
from trying to do this; instead, one further issue is as displayed
in the next section.

\newsec{Holography}

Black holes are not only legitimate solutions of Einstein's field
equations for the gravitational force, one can also show quite
easily that black holes \emph{inevitably} form under given
favorable initial conditions of conventional matter
configurations. Such `conventional' black holes are very big,
having a radius at least of the order of 10 km. Therefore, they
are usually considered as classical, \emph{i.e.} non-quantum
mechanical, objects. But, at least in principle, they should also
obey the laws of Quantum Mechanics. Elementary particles in the
vicinity of a black hole should be described by Quantum Field
theory, and the laws of General Relativity should dictate how to
handle Quantum Field theory here. As was shown by S.~Hawking, this
exercise leads to the astonishing result that particles must
emerge from a black hole.\cite{HH}

Mathematically, the explanation for this effect is that
\emph{time} is measured by freely falling observers in a
coordinate frame that is fundamentally different from the
coordinate frame used by the onlooking observer outside the black
hole.  Physically, one may explain the emission as a
gravitational tunneling effect, comparable to the pair creation
of charged particles in the presence of a strong electric field.
The emission rate is precisely computable, and conventional theory
gives a flux of particles corresponding to a temperature \be
T_{\rm H}={\hbar c^3\over 8\pi k\, G\, M_{\rm BH}}\
,\le{THawking} where \(k\) is Boltzmann's constant, and \( M_{\rm
BH}\) is the mass of the black hole.

Hawking's result can be used to estimate the \emph{density of
quantum states} of a black hole. Assuming a transition amplitude
\(\mathcal{T}_{\rm in}\) for the absorption process, we can write
in two ways an estimate for the absorption cross section
\(\s(\vec k)\) for an amount of matter \(\d E\) with momentum
\(\vec k\) by a black hole of mass \(M_{\rm BH}\):
\be \s\approx&2\pi r_+^2=8\pi M_{\mathrm BH}^2\,;\label{absorption}\\
 \s=&|\mathcal{T}_{\mathrm in}|^2\r(M_{\mathrm BH}+\d
E)/v\,.\le{sigma} Here, \(r_+\) is the radius of the outer event
horizon, \(\r(M)\) is the density of states of a black hole with
mass \(M\), and \(v\) is the velocity of the absorbed particle.
The probability \(W\,{\mathrm d}t\) of a particle emission during
a time interval \({\mathrm d}t\) can also be written in two ways:
\be  W\,\mathrm d t =&|\mathcal{T}_{\mathrm out}|^2\,
\r(M_\mathrm{BH})\,\mathrm d t/V\,;\\ W\,{\mathrm d}t  =
&\displaystyle{\s(\vec k)v\over V}\,e^{-\d E/kT_\mathrm H} \mathrm
d t\,.\le{emission} Here, \(V\) is the volume of a box, in which
the wave function of the emitted particle is normalized. Dividing
Eqs. \eqn{absorption} --- \eqn{emission}, we get, in Planck
units,\cite{gthBH} \be {\r(M+\d
E)\over\r(M)}={|\mathcal{T}_{\mathrm out}|^2\over
|\mathcal{T}_{\mathrm in}|^2}\,e^{\d E/kT_\mathrm H}=e^{8\pi M\,\d
E}\ .\le{ratio}

We assumed here that \(|\mathcal{T}_{\mathrm
out}|=|\mathcal{T}_{\mathrm in}|\). All that is needed for this
assumption is \(PCT\) invariance, since \(\s(\vec k)\) is
symmetric under \(P\) and \(C\). For all known field theories,
\(PCT\) is a perfect symmetry. Needless to say, we do not know
this to be so for quantum gravity, but it would be a natural
assumption.

Eq.~\eqn{ratio} is to be seen as a differential equation that is
easily integrated, to give: \be\r(M)=e^{4\pi M^2+\mathcal
C}=\mathcal C'\,2^{A/A_0}\,,\le{states} where \(\mathcal C\) and
\(\mathcal C'\) are integration constants, \(A=4\pi r_+^2\) is the
black hole area, and \(A_0 =4\ln2\) in Planck units. One concludes
that the density of quantum states of a black hole is that of an
object with \(A/A_0\) free Boolian parameters on its surface. The
integration constant represents a fixed degree of freedom that all
black holes have in common. The result \eqn{states} can also be
derived using thermodynamics, but then one has to cope with the
difficulty that black holes, embedded in a thermal environment,
are unstable because of their negative specific heat (they cool
off when energy is added to them).\cite{thermo}

In one respect, this result appears to be quite interesting and
acceptable. Apparently, the quantum states of a black hole form a
discrete set, just as if the black hole were a fairly ordinary
object, easily to become macroscopic, in the astronomical case.
Black hole formation and evaporation can indeed be described in
terms of quantum amplitudes, and if the black hole is very tiny,
these amplitudes can be represented in Feynman diagrams.

On closer inspection, however, there are several problems with
this result. We would have thought that a general coordinate
transformation transforms \emph{states} into \emph{states}. An
in-going observer describes what (s)he sees in terms of particle
states superimposed on an approximately flat space-time
environment, using regular coordinates. The outside observer uses
the black hole coordinates featuring an horizon. The states
observed by the outside observer are counted by discrete
variables of one bit for every area unit \(A_0\). How can this
mapping of discrete states onto the continuum of states for the
in-going observer be unitary? How does the in-going observer
count his/her states? We should have expected the number of these
states to scale with the bulk volume of his environment, not with
the area.

Hawking's calculation gives no clues here. To the contrary, it
appears to tell us that, even if the initial state of an imploding
object would be a quantum mechanically pure state, the radiating
black hole that emerges after some time would nevertheless be in a
quantum mechanically mixed state.\cite{hawkmix} Such a transition
cannot be described by any Schr\"odinger equation. Does the black
hole, viewed as an isolated object, disobey the quantum code? This
is what was concluded initially, but most of us now agree that
such a conclusion must be premature.\cite{gthBH}

If, however, on the other hand, information is conserved in
unitary evolution equations, how is it that the information in the
in-going particles is transmitted to the out-going ones?

A first clue towards answering this question was provided by taken
into account the fact that ingoing particles interact with the
outgoing ones when they pass each other. The gravitational
interaction here diverges. Early ingoing particles meet late
outgoing ones in an entirely different local Lorentz frame, so
that the relative energy, that is, the energy in the center of
mass frame, is large, and this number diverges exponentially with
the time difference of the two particles.\cite{gthstrings} Taking
this into account, one does find a unitary scattering matrix
relating outgoing particles to the ingoing ones, but the spectrum
of states does not seem to be bounded by the horizon area. Such a
bound presumably has to come from the transverse components of the
gravitational interactions, which is much harder to
calculate.\cite{SdH}

Requiring that the number of states in some region of space,
described by a theory, is bounded by the surface area of this
region, seems to be paradoxical. This paradox seems to be as deep
and fundamental as the one that lead M.~Planck to his postulates
of Quantum Mechanics, or, in other words, we expect that its
resolution requires a paradigm shift. How can we have locality in
three-space, but numbers bound by two-space?

In certain versions of string theories, these apparently
conflicting demands are met to some extent,\cite{JM} except that
the concept of locality appears to be ignored. The amount of
`magic' required for these ideas to work is still not acceptable.
It is this author's belief that the true reason for the mysterious
nature of this problem is our insistence to stick to the language
of Quantum Mechanics. It seems to be only natural to see a link
between the mysteries of string theory and those of the correct
interpretation of Quantum Mechanics.

\newsec{Harmonic oscillators.}
It is instructive to ask how a deterministic system can be
addressed using the mathematics of Quantum Mechanics. Our starting
point is that we may have simple autonomous dynamical systems,
where later we will decide how they should be coupled. Thus, we
start with a deterministic system consisting of a set of \(N\)
states,\cite{disdet} \be(0),(1),\cdots,(N-1)\}\nm\ee on a circle.
Time is discrete, the unit time steps having length \(\t\) (the
continuum limit is left for later). The evolution law is:
    \be t\ra t+\t\quad:\qquad (\n)\ra(\n+1\,{\rm\ mod}\,
    N)\,.\le{eq2.1}
Introducing a basis for a Hilbert space spanned by the states
\((\n)\), the evolution operator can be written as
    \be U(\D t=\t)\ =\ e^{-iH\t}\ =e^{-\textstyle{\pi i\over
    N\vphantom{_g}}}\pmatrix{0&&&&1\cr 1\,&0\cr &1&0\cr
    &&\ddots&\ddots\cr&&&1&0\cr }\ .\le{eq2.2}
The phase factor in front of the matrix is of little importance;
it is there for future convenience. Its eigenstates are denoted as
\(|n\ket\), \(n=0,\cdots,N-1\). They are found to be \be
|n\ket={1\over\sqrt N}\sum_{\n=1}^N e^{2\pi i n\n\over N}(\n)\
,\quad n=0,\cdots,N-1\ .\le{Estates}

This law can be represented by a Hamiltonian using the notation
of quantum physics:
    \be H|n\ket={2\pi(n+\half)\over N\t}|n\ket\,.\le{eq2.3}
The \(\half\) comes from the aforementioned phase factor. Next, we
apply the algebra of the \(SU(2)\) generators \(L_x\), \(L_y\) and
\(L_z\), so we write
    \be N\deff 2\ell+1\quad,\qquad n\deff m+\ell\quad,\qquad
    m=-\ell,\cdots,\ell\ .\le{eq2.4}
Using the quantum numbers $m$ rather than $n$ to denote the
eigenstates, we have
    \be H|m\ket={2\pi(m+\ell+\half )\over (2\ell+1)\t}|m\ket\qquad\hbox{or}\qquad
    H={2\pi\over (2\ell+1)\t} \,(L_z+\ell+\half)\ .
    \le{eq2.5}
This Hamiltonian resembles the harmonic oscillator Hamiltonian
when \hbox{$\w=2\pi/(2\ell+1)\t$}, except for the fact that there
is an upper bound for the energy. This upper bound disappears in
the continuum limit, if \(\ell\ra\infty\), \(\t\downarrow 0\).
Using \(L_x\) and \(L_y\), we can make the correspondence more
explicit. Write
    \be L_\pm|m\ket&\deff \sqrt{\ell(\ell+1)-m(m\pm1)}\ |m\pm1\ket\
    ;\label{eq2.6}\\  L_\pm&\deff L_x\pm iL_y \quad;\qquad[L_i,L_j]=i\e_{ijk}L_k\
    , \le{eq2.6b}
and define
    \be\hat x\deff\a L_x\quad,\qquad \hat p\deff\b L_y\quad;\qquad
    \a\deff\sqrt{\t\over\pi}\quad,\qquad\b\deff{-2\over
    2\ell+1}\sqrt{\pi\over\t} \ .\le{eq2.7}
The commutation rules are
    \be[\hat x,\hat p]=\a\b
    iL_z=i(1-{\t\over\pi}H)\,,\le{eq2.8}
and since
    \be L_x^2+L_y^2+L_z^2=\ell(\ell+1)\,,\le{eq2.9} we have
    \be H=\half\w^2 \hat x^2+\half \hat p^2+ {\t\over2\pi}\left({\w^2\over 4}+
    H^2\right)\,. \le{eq2.10}

Now consider the continuum limit, \(\t\downarrow 0\), with
\(\w=2\pi/(2\ell+1)\t\) fixed, for those states for which the
energy stays limited. We see that the commutation rule
\eqn{eq2.8} for \(\hat x\) and \(\hat p\) becomes the
conventional one, and the Hamiltonian becomes that of the
conventional harmonic oscillator: \be [\hat x,\hat p]\ra i\
;\qquad H\ra \half\w^2 \hat x^2+\half \hat p^2\ .\le{contlimit}

There are no other states than the legal ones, and their energies
are bounded, as can be seen not only from \eqn{eq2.10} but rather
from the original definition \eqn{eq2.5}. Note that, in the
continuum limit, both \(\hat x\) and \(\hat p\) become continuous
operators, since both \(\a\) and \(\b\) tend to zero.

The way in which these operators act on the `primordial' or
`ontological' states \((\n)\) of Eq.~\eqn{eq2.1} can be derived
from \eqn{eq2.6} and \eqn{eq2.7}, if we realize that the states
\(|m\ket\) are just the discrete Fourier transforms of the states
\((\n)\), see Eq.~\eqn{Estates}. This way, also the relation
between the eigenstates of \(\hat x\) and \(\hat p\) and the
states \((\n)\) can be determined. Only in a fairly crude way,
\(\hat x\) and \(\hat p\) give information on where on the circle
our ontological object is; both \(\hat x\) and \(\hat p\) narrow
down the value of \(\n\) of our states \((\n)\).

The  most important conclusion from this section is that there is
a close relationship between the quantum harmonic oscillator and
the classical particle moving along a circle. The period of the
oscillator is equal to the period of the trajectory along the
circle. We started our considerations by having time discrete, and
only a finite number of states. This is because the continuum
limit is a rather delicate one. One cannot directly start with the
continuum because then the Hamiltonian does not seem to be bounded
from below.

The price we pay for a properly bounded Hamiltonian is the square
root in Eq.~\eqn{eq2.6}; it may cause complications when we
attempt to introduce interactions, a problem that is not yet
properly worked out.

Starting from this description of harmonic oscillators in terms of
deterministic models, one may attempt to construct deterministic
theories describing, for instance, free bosonic particles, see
Ref.\cite{gthboson}

Strings can also be seen as collections of harmonic oscillators.
A first attempt to write string theory in deterministic terms
failed because conformal invariance could not be built
in.\cite{detstr} Apparently, further new ideas are needed here.

\newsec{Continuous degrees of freedom.}

In the previous section, a discrete, periodic system was
considered, and we took the continuum limit in the end. Could one
not have started with a continuous model right from the beginning?

Take a Newtonian equation, \be {\dd\over\dd t}q^i(t)=f^i(\vec
q)\,.\le{contevol} We can write the \emph{quantum} Hamiltonian,
\be H=\sum_i p_if^i(\vec q)\ ,\qquad p_i={\hbar\over
i}{\pa\over\pa q^i}\ .\le{hamilton} This is quantum language for a
classical, deterministic system. It works because the Hamiltonian
is linear, not quadratic, in the momenta \(p_i\). The difficulty
linking this with real Quantum Mechanics is that this Hamiltonian
cannot possibly be bounded from below, so that there is no ground
state.

\newsec{Massless, noninteracting fermions.}

Massless, non-interacting fermions are entirely deterministic.
this can be demonstrated by identifying the `beables' for this
system. Beables are a complete set of observables \(\OO_i(t)\)
that commute at all times: \be [\OO_i(t),\ \OO_j(t')]=0\ ,\quad
\forall\ t,\,t'\,.\le{beables} First, consider only
first-quantized, chiral fermions. They have a two-component
complex wave function obeying the Hamilton equation for \be
H=\vec\s\cdot\vec p\ ;\qquad
\s_i\,\s_j=\d_{ij}{\Bbb{I}}+i\e_{ijk}\s_k\ ,\le{chiralH} \(\s_i\)
being the Pauli matrices. Consider the set \be\OO_i(t)=\{\hat p,\
\hat p\cdot\vec\s,\ \hat p\cdot\vec x\}\ ,\le{neutrinobeables}
where \be \hat p_i\equiv\pm{p_i\over|p|}\ ,\qquad \hat
p_x>0\,.\le{phat} These operators obey closed time evolution
equations: \be{\dd\over\dd t}\vec x=\vec\s(t)\ ,\qquad{\dd\over\dd
t}\left(\hat p\cdot\vec x(t)\right)=\hat p\cdot\vec\s\ , \qquad
{\dd\over\dd t}\left(\hat p\cdot\vec\s\right)=0\ .\qquad\qquad{
}\\ \hat p(t)=\hat p(0)\ ,\qquad \hat p\cdot\vec\s(t)=\hat
p\cdot\vec\s(0)\ ,\qquad \hat p\cdot\vec x(t)=\hat p\cdot\vec
x(0)+\hat p\cdot\s(0)t\ . \le{eqfermions} The fact that all
operators in Eq.\eqn{neutrinobeables} commute with one another is
easy to establish, with the possible exception of \([\hat
p_i,\,\left(\hat p\cdot\vec x\right)]\). The fact that the latter
vanishes is most easily established in momentum space, realizing
that \(\vec p\cdot\vec x\) is the dilatation operator, while
\(\hat p\) keeps the same length 1 under dilatations: \be[(\hat
p\cdot\vec x),\,\hat p_i]=i\left(\hat p\cdot{\pa\over\pa\vec
p}\right) \hat p_i=0\,.\le{derivation}

The \emph{physical} interpretation of this result is that the
dynamical behaviour of a massless, chiral, non-interacting fermion
is exactly like that of an \emph{infinite, flat, oriented sheet},
moving with the speed of light in a direction orthogonal to the
sheet. \(\pm\hat p\) gives the direction of the sheet, \(\hat
p\cdot\vec\s\) gives the sign of its orientation and \(\hat
p\cdot\vec x\) its distance from the origin.

As before, we encounter the difficulty that, in this
deterministic system, the Hamiltonian is not bounded below, again
because it is linear in the momenta \(p_i\). Thus, there exists
no ground state. In this case, however, P.A.M.~Dirac told us what
to do: \emph{second quantization}. Assume an indefinite number of
particles with hamiltonian \eqn{chiralH}. Consider the range of
energies they can have. Take the state where all negative energy
states are occupied by a particle, all positive energy states are
empty. That is the state with lowest possible total energy, the
vacuum state. It is standard procedure, but it does require our
particles to obey Fermi statistics, or, in other words, no two
particles are allowed in the same state, and interchanging two
particles does not change a state into a different one.

The latter condition is easily met, but to forbid two particles
to be in the same state requires some sort of repulsion. The
easiest procedure is to have at each value for the unit
orientation vector \(\hat p\) a grid with some finite spacing
\(a\). No two sheets are allowed on the same lattice point. Then
we can count the states exactly as in a fermion theory and the
second quantization procedure works. The limit \(a\downarrow 0\)
can be taken without any difficulty.

\newsec{Locality.} Let us focus a bit more on the ontological
states for massless fermions. They are characterized by an
orientation \(\hat k\) (obeying \(|\hat k|=1\)), and a distance
scalar \(z\). Furthermore, we need the quantum operator
conjugated to \(z\), which we call \(q\equiv -i\pa/\pa z\). We
define \(\et_3\) to be the sign of \(q\), and \(\hat k\) is the
orientation of the sheet with its sign chosen such that it moves
in the positive \(\hat k\) direction: \be(\hat p\cdot\vec\s)\hat
p\ \equiv\ \hat k\ \equiv\ \et_3{\vec p\over|p|} \ ;\qquad
\et_3\equiv{(\vec p\cdot\vec \s)\over |p|}\ =\ \pm1\ . \le{khat}

\(z=(\hat k\cdot\vec x)\) is the distance of the sheet from the
origin, apart from its sign, which denotes whether the sheet moves
away from or towards the origin. To define the original components
of the vector \(\vec p\), we first have to find its length
\(|p|\). This we take to be the operator \be|p|\equiv |q|\equiv
-i\et_3{\pa\over\pa z}\ ,\qquad\hbox{so}\qquad \vec p=-i\hat
k\,{\pa\over\pa z}\ . \le{momentum}

\(\hat k\) and \(z\) are the ontological variables, or beables,
whereas \(q\) and \(\et_3\) are changeables. We have \(H=q\), so
the dynamical equations are now simply \be \dot z=1\ ;\qquad
\dot{\hat k}=0\ ,\le{deteq} which are the equations of a sheet
moving in a fixed direction. Since Eq.~\eqn{momentum} defines the
momenta, and their canonically conjugated operators the
positions, we should now be in a position to compute the
conventional wave function \(\j(\vec x,\s_3)\), \(\s_3=\pm 1\), if
we have some wave function \(\j(\hat k,z)\). A fairly delicate
calculation gives \be \j( \vec
x,\pm)={1\over2\pi}\vec\pa_x^{\,2}\int\sin\th\,\dd\th\,\dd\vv
\pmatrix{ \cos\half\th\cr e^{-i\vv}\sin\half\th}|\hat k,z\ket\
,\le{wave} where we used the notation \be z=(\hat k\cdot\vec x)\
;\qquad \hat
k=\left\{\matrix{\sin\th\cos\vv\cr\sin\th\sin\vv\cr\cos\th}
\right\} \,.\le{zkhat} Thus, apart from the Laplacian, all sheets
contributing to \(\j(\vec x,\s_3)\) are going through the point
\(\vec x\).

\begin{figure}
\epsfxsize=80 mm\epsfbox{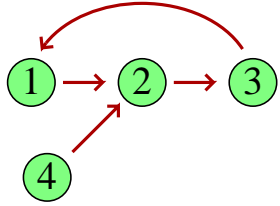} 
  \caption{\footnotesize{Mini-universe with information loss.
  The arrows show the evolution law.}}\label{fig:infolossmod}
\end{figure}

\newsec{Information loss}
The reasons why \emph{information loss} may be an essential
ingredient in deterministic hidden variable models of the sort
pioneered above, has been extensively discussed in
Ref.\cite{disdet},\cite{BJV}. A prototype microcosmos with
information loss is the model of Fig.~\ref{fig:infolossmod}.
Following the arrows, one would conclude that the evolution matrix
is \be U=\pmatrix{0&0&1&0\cr 1&0&0&1\cr 0&1&0&0\cr
0&0&0&0}\,.\le{evolveloss} This, of course, is not a unitary
matrix. One way to restore unitarity would be to remove state \#
4. The problem with that is that, in universes with tremendously
many allowed states, it would be very difficult to determine which
of the states are like state number 4, that is, they have no state
at all in their (distant) past.

A preferred way to proceed is therefore to introduce
\emph{equivalence classes} of states. Two states are equivalent
iff, some time in the near future, they evolve into one and the
same state\fn{It could also happen that two states merge into the
same state in the \emph{distant} future, but in many models such
events become increasingly unlikely as time goes on.}. In
Fig.~\ref{fig:infolossmod}, states \#\# 1 and 4 are equivalent,
so they form one class. By construction then, equivalence classes
evolve uniquely into equivalence classes.

It should be emphasized that, at the Planck scale, information
loss is not a small effect but a very large effect. Large numbers
of `ontological' states are in the same equivalence class, and
the equivalence classes form a much smaller set than the class of
all states. This is how it can happen that the total number of
distinguishable quantum states (= the number of equivalence
classes) may only grow exponentially with the \emph{surface} of a
system, whereas the total number of ontological states may rise
exponentially with the volume. This seems to be demanded by black
hole physics, when we confront the laws of quantum mechanics with
those of black holes.

Information loss at the level of the underlying deterministic
theory, may also explain the apparent lack of causality in the
usual attempts to understand quantum mechanics in terms of hidden
variables. The definition of an equivalence class refers to the
future evolution of a system, and therefore it should not be
surprising that in many hidden variable models, causality seems
to be violated. One has to check how a system will evolve, which
requires advance knowledge of the future.

Information loss at the Planck scale may also shed further light
on the origin of gauge theories. it could be that, at the level
of the ontological degrees of freedom at the Planck scale, there
is no local gauge symmetry bat all, but in order to describe a
physical state, that is, an equivalence class, we need to describe
a particular member of this class, a single state. its relation
to the other members of the same equivalence classes could be
what is presently called a `gauge transformation".

There is another aspect to be considered in theories with
information loss. Theories with continuous degrees of freedom
would have an infinity of possible states if there were no
information loss. With information loss, there may be a discrete
set of limit cycles, meaning that the equivalence classes may
still form discrete sets. Discreteness, one of the prime
characters of quantum physics, could thus be ascribed to
information loss.

\newsec{Conclusions.} Our view towards the quantum mechanical
nature of our world can be summarized as follows.\begin{itemize}
\item Nature's fundamental laws are defined at the Planck scale.
At that scale, all we have is \emph{bits of information}.
\item A large fraction of this information gets lost very quickly,
but it is being replenished by information entering from the
boundaries.
\item A quantum state is defined to be an equivalence class of
states which all have the same distant future. This definition is
non-local and acausal, which implies that, if we would attempt to
describe everything that happens purely in conventional quantum
mechanical terms, such as what is done in superstring theories,
locality and even causality will seem to be absent at the Planck
scale. Only in terms of a deterministic theory this demand of
internal logic can be met.
\item These equivalence classes are described by observables that
we call `beables'. In quantum terminology, beables are a complete
set of operators that commute at all times, see Eq,\eqn{beables}.
A beable describes what a Planckian observer would be able to
register about a system --- information that did not get lost.
\item All other quantum operators are `changeables', operators
that do not commute with all beables.
\item The wave function \(\j\) has the usual Copenhagen/Bohr/Bohm
interpretation, but:
\item Many or all of the familiar symmetries of Nature, such as
translation, rotation, Lorentz and isospin symmetry, must be
symmetries that relate beables to changeables. This means that
the `ontological' theory behind Quantum Mechanics does not have
these symmetries in a conventional form. \end{itemize} When we go
from the Planck scale to the Standard Model scale,\begin{itemize}
\item Our only way to obtain effective laws of physics at the
larger distance scales is by applying the renormalization group
procedure.
\item Beables and changeables are then mixed up to the extent that
it is impossible to identify them; they obey the same laws of
physics.
\item When we perform a typical quantum experiment, we therefore
do not know in advance whether an operator we are working with is
a beable or a changeable. Due to the symmetries mentioned above,
beables and changeables may obey the same laws of physics. Only
when we measure something, and not before, do we know that what we
are looking at is a beable. In this way, we believe, apparent
clashes with Bell's inequalities may be avoided.
\item The classical observables in the classical (macroscopic)
limit, commute with the beables. They are beables as well.
\end{itemize}
There are numerous difficulties left. Most urgent is the need for
a viable model, demonstrating how the mechanism works that we
believe to be responsible for the conspicuous quantum mechanical
nature of the world that we live in. It continues to be difficult
to produce a non-trivial model, one showing particles that
interact, for instance, such that its Hamiltonian is bounded from
below.

\end{document}